\documentclass{iau_FM}
\usepackage{graphicx}
\usepackage{natbib}

\title[FM5.~~Exo-space weather] 
{Stellar space weather effects on potentially habitable planets}

\author[Vidotto]   
{A.~A.~Vidotto}

\affiliation{Leiden Observatory, Leiden University, PO Box 9513, 2300 RA Leiden, The Netherlands}

\pubyear{2022}
\jname{Astronomy in Focus, Volume 1} 
\editors{Piero Benvenuti, ed.}
\begin{document}

\maketitle

\begin{abstract}
Stellar activity can reveal itself in the form of radiation (eg, enhanced X-ray coronal emission, flares) and particles (eg, winds, coronal mass ejections). Together, these phenomena shape the space weather around (exo)planets. As stars evolve, so do their different forms of activity -- in general, younger solar-like stars have stronger winds, enhanced flare occurrence and likely more frequent coronal mass ejections. Altogether, these effects can create harsher particle and radiation environments for habitable-zone planets, in comparison to Earth, in particular at young ages. In this article, I will review some effects of these harsher environments on potentially habitable exoplanets. 
\keywords{MHD, planets and satellites: atmospheres, magnetic fields, planet-star interactions}
\end{abstract}

\firstsection 
\section{Introduction}
The short review is based on the review talk I gave at the ``Focus Meeting 5: Beyond the Goldilocks Zone: the Effect of Stellar Magnetic Activity on Exoplanet Habitability'' (August 2022). In line with the theme of the meeting, my review talk focused in particular on the effects of stellar space weather on planets that could be potentially habitable. Therefore, two important concepts are necessary here: what do I mean by potentially habitable planet? and what do I mean by space weather? 

In the context of this article, I consider a potentially habitable planet as a planet that orbits within the habitable zone. The habitable zone is the region around a star wherein an orbiting planet can sustain temperatures that allow water to remain liquid at its surface. Because the planetary equilibrium temperature depends on the irradiation the planet receives from its host star, and because the incoming radiation flux depends on orbital distance ($a_{\rm orb}$) and luminosity ($L_\star$) as $\sim L_\star a_{\rm orb}^{-2}$, this means that an Earth twin needs to orbit closer to a less luminous (cooler) main-sequence star to maintain water in liquid form at its surface. In the main-sequence phase, the luminosity of the star depends on its mass as $L_\star \propto M_\star^\alpha$, with $\alpha \simeq 4$. Therefore, this implies that the distance an Earth twin would have to orbit to remain at the same temperature as Earth would be $a_{\rm orb} (\textrm{au}) \simeq (L_\star/L_\odot)^{-1/2} \simeq (M_\star/M_\odot)^2$.\footnote{This is an rough derivation of the habitable zone distance as a function of the stellar mass. More rigorous derivations can be found in works that more completely consider atmospheric properties of  planets and mass-luminosity relations from stellar evolution models \citep[e.g.][]{2007A&A...476.1373S,2014ApJ...787L..29K}.}

Regarding the concept of exo-space weather, here I consider it to be broadly related to certain stellar phenomena (``activity'') that shape the environment around planets. Stellar activity can reveal itself in the form of radiation (e.g.,  X-ray coronal emission, flares) and particles (e.g., winds, coronal mass ejections). In cool main-sequence stars, stellar high-energy radiation (X-ray and EUV fluxes) decreases with age \citep[e.g.,][]{2005ApJ...622..680R,2015A&A...577L...3T}. Likewise, stellar particle fluxes (or mass-loss rates) also decay with age \citep{2021ApJ...915...37W, 2021LRSP...18....3V}. Together, these two factors indicate that planets orbiting around more active (which are generally younger) stars would experience stronger space weather effects compared to planets in the present-day solar system. Interestingly, similarly harsher environments are also experienced by planets orbiting older stars at close-in orbits \citep{2015MNRAS.449.4117V}. Therefore, studying the exo-space weather of close-in planets orbiting less active, older stars could give us important clues as to the space weather effects experienced by the (much farther out) planets in the solar system at younger ages.

\section{Exo-space weather and planetary magnetic fields} 
One important ingredient for habitability is the presence of a planetary atmosphere. For unmagnetised planets, stellar wind particles can directly interact with the atmosphere of the planet -- this could strip away the atmosphere. For magnetised planets, the situation is less clear. On one hand, the planetary magnetic field shields (at least, part of) the atmosphere from the direct interaction with stellar wind particles. In the case of the Earth, for example, the stand-off distance between the interaction zone with the stellar wind and the planet surface can be found by equating the ram pressure (or the dynamic pressure) of the local solar wind and Earth's magnetic pressure \citep{1930Natur.126..129C}: $P_{\textrm{ram}} (\textrm{wind}) = P_B (\textrm{planet})$.  With the assumption that the magnetic field is well described by a dipole, then $P_B \simeq  B_p^2 (R_p/r)^6 / (8\pi)$, with $B_p$ being the planetary magnetic field at the equator of the planet, $R_p$ the planetary radius and $r$ the distance to the centre of the planet. Therefore, one can derive the distance where this interaction takes place to be $r_M/R_p = \sqrt{2}B_p^{1/3} /(8\pi P_{\textrm{ram}} (\textrm{wind}) )^{1/6}$\footnote{A factor of $\sqrt{2}$ is often included in the Chapman-Ferraro equation to match Earth's observation.}. In the case of the Earth, the ram pressure of the local solar wind is $P_{\textrm{ram}} \simeq m_p n_{\rm sw} u_{\rm sw}^2 = m_p (5 ~{\rm particles/cm}^3) \times (400 {\rm ~km/s})^2 \simeq 1.3 \times 10^{-8} {\rm dyn/cm}^2$, where the number density and velocity of the local solar wind (i.e., at Earth's orbit) is given by $n_{\rm sw}$ and  $u_{\rm sw}$, respectively. With an equatorial magnetic field strength of about 0.3~G,  we find a distance of about $\sim 11 R_\oplus$ \citep{2013pss3.book..251B} for the size of Earth's magnetosphere, which is much greater than the thickness of Earth's atmosphere. 

\begin{figure}[t]
\begin{center}
 \includegraphics[width=0.6\textwidth]{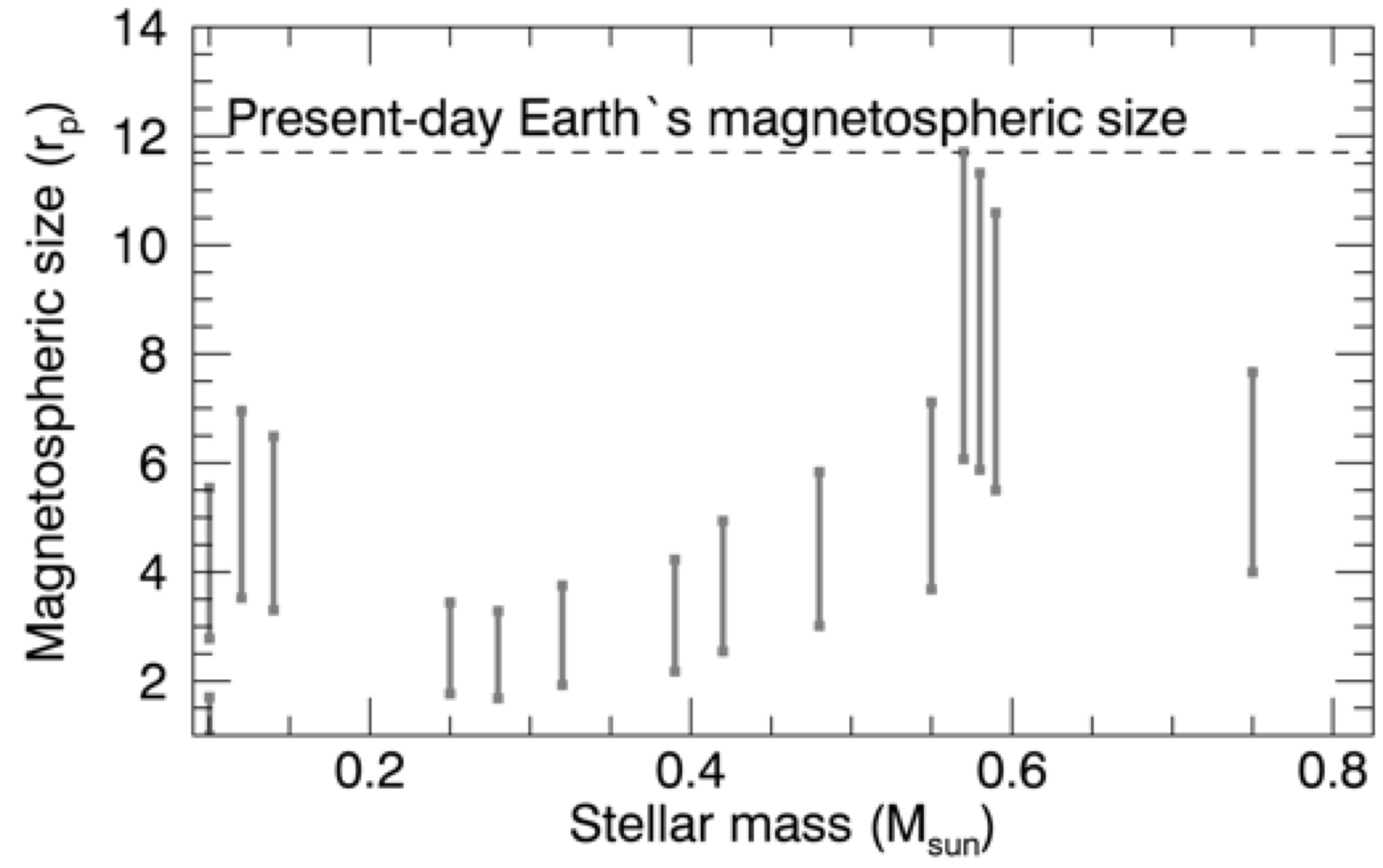} 
\end{center}
 \caption{Earth-analog at habitable zones of M dwarfs: Intense stellar magnetism can reduce the extent of planetary magnetospheres. Figure based on \citet{2013A&A...557A..67V}.} \label{fig1}
\end{figure}

More generally, the pressure balance equation should take into account not only the local ram pressure of the stellar wind, but also other pressure terms (e.g., the magnetic and thermal pressures). In practice, for the planets in the solar system, these other terms are usually much smaller than the ram pressure and thus are neglected. However, for a planet orbiting a more active star (with stronger magnetism), the other terms should also be incorporated. In \citet{2013A&A...557A..67V}, I demonstrated that planets with a similar magnetisation as that of the Earth, if orbiting in the habitable zone of active M dwarfs, could have magnetospheres  substantially smaller than that of the Earth (see Figure \ref{fig1}). In one case, the fictitious (potentially) habitable planet orbing WX UMa would have its magnetosphere completely crushed \citep[WX UMa has a very strong magnetic field \citealt{2010MNRAS.407.2269M} and thus a harsh wind environment as well, as presented in ][]{2022MNRAS.514..675K}. \citet{2013A&A...557A..67V} concluded that, even when neglecting the pressure of the stellar wind, Earth-like planets around M dwarfs might not have Earth-like magnetospheres. In \citet{2014A&A...570A..99S}, it was shown that for Earth-like planets in the habitable zone of K, G and F stars with different activity levels, the situation is less extreme. As the star ages, its activity and wind strength decrease, and thus magnetospheres are allowed to ``expand'' \citep{2018MNRAS.476.2465O, 2019MNRAS.489.5784C}, which in theory means that they would better shield (at least, in part)  the atmospheres of potentially habitable planets.

Although the planetary magnetic ``bubble'' can prevent direct interaction between the stellar wind and the entire atmosphere of the planet, its protective effect has been questioned recently \citep{2013cctp.book..487B}. The magnetosphere (assuming a dipolar field) consists of regions of closed field lines and regions of open field lines. The open field line regions do not provide a shield against stellar wind particles -- in fact, an inflowing stellar wind can be channeled towards the polar regions through open magnetic field lines. This could then lead to increased local heating which would enhance atmospheric evaporation through polar/auroral flows. This then leads to the question whether a magnetic field could lead to retention of planetary atmosphere (the shield scenario) or could enhance atmospheric loss (the polar flow scenario). 

Maybe the answer is not a clear cut distinction between both arguments, but rather a combination of them.  \citet{2018MNRAS.481.5146B} suggested that, throughout the life of the Earth, there has been a competition between  two processes: an increased collection area for solar wind mass capture and reduced solar wind flux, but that overall the Earth's magnetic field has provided a net protective role throughout Earth's history.

\section{Effects of high-energy radiation on planetary atmospheres}
\begin{figure}[t]
\begin{center}
 \includegraphics[width=0.5\textwidth]{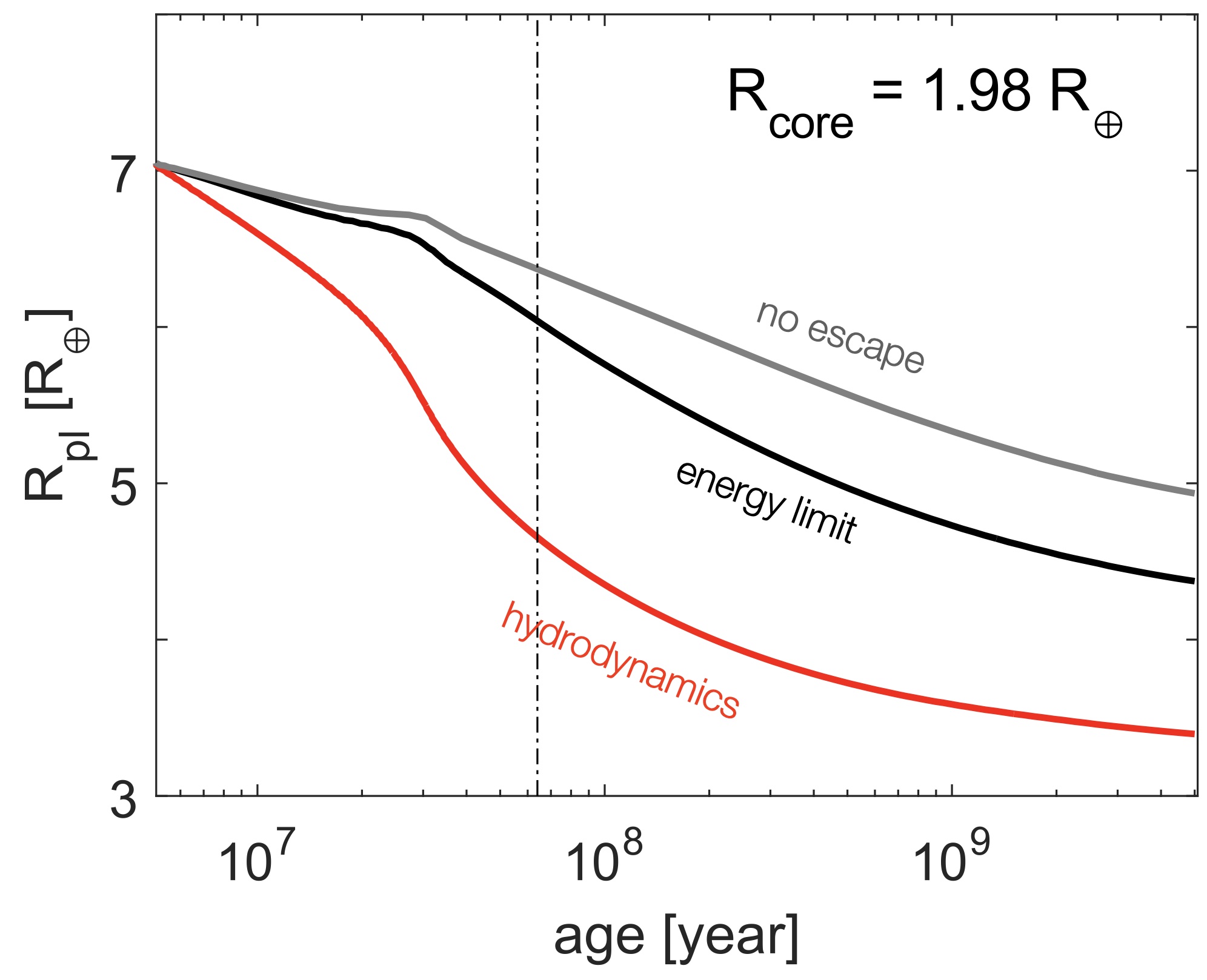} 
\end{center}
 \caption{Evolutionary tracks of the radius of  a close-in terrestrial planet, considering different prescriptions for atmospheric evaporation hydrodynamic model (red line) and energy limited approximation (black line). Note that in the absence of atmospheric escape (grey line), the radius of the planet changes solely due to thermal evolution. Figure based on \citet{2020MNRAS.499...77K}. Planetary evolution with atmospheric escape is computed using MESA and the inlists are publicly available in \citet{zenodo_daria_mesa}.} \label{fig2}
\end{figure}

In cool main-sequence stars, particle fluxes (winds, coronal mass ejections) and high-energy radiation are ultimately connected to stellar magnetism: magnetic fields drive stellar winds and also heat stellar chromosphere and coronae \citep{2019ARA&A..57..157C}. Therefore, it is not surprising that both stellar wind particle fluxes and high-energy radiation are higher in younger stars and decrease as stars age \citep{2021LRSP...18....3V}. The EUV flux that a planet receives plays, in particular, a major role in atmospheric escape. There are three timescales involved in the process. 
\begin{enumerate}
\item From an evolutionary point of view ($\sim$ several Gyr), the decrease in high-energy radiation from the host star implies that its effect on atmospheric evaporation would also decrease with time. \citet{2015A&A...577L...3T} demonstrated that the hydrogen content of the atmosphere of a terrestrial planet orbiting at 1~au of its solar-like stellar host can evolve significantly through the evolution of the system. They showed that a planet could even lose its entire primordial atmosphere very rapidly. This happens, for instance, if the host star started its life with a higher rotation -- in this case, the host star remains more active (i.e., with a higher EUV luminosity) for the first billion year of its life than a star that started its life with a slower rotation (fast rotating stars are more active than slowly rotating stars, e.g., \citealt{2003A&A...397..147P}). Consequently, the survival of planetary atmospheres depends on the history of the high-energy radiation of the host star \citep{2015A&A...577L...3T, 2019MNRAS.490.3760A, 2021A&A...654L...5P}. Another interesting point to notice is that, when substantial material is lost through atmospheric evaporation, the internal structure of the planet readjusts itself \citep[e.g.,][]{2020MNRAS.499...77K, 2021MNRAS.504.2034K}, implying that the evolutionary track of a planet also depends on the history of atmospheric escape (see Figure \ref{fig2}).\footnote{This is not too dissimilar to what happens in the evolution of massive stars, whereby  stellar winds can carry away a substantial fraction of the stellar mass, changing its internal structure, the amount of energy it produces, and consequently its evolutionary track in the HR diagram.}
\item The high-energy radiation of solar-type stars  also evolves during activity cycles. For example, in the case of the Sun, both its X-ray and EUV luminosities evolve every $\sim 11$ years \citep{2000ApJ...528..537P,2016A&A...587A..87C}, with an evolution that follows closely the trend of sunspot number \citep{2020MNRAS.496.4017H}. As a consequence, as atmospheric escape is influenced by the high-energy input radiation, escape process should also evolve within activity cycle timescales.
\item On a shorter timescale,  on the order of an hour or so, escape can also change with increase in energy input caused by flares \citep[e.g.,][]{2020MNRAS.496.4017H}. Additionally, if flares are accompanied by stellar coronal mass ejections, the associated particle flux of coronal mass ejections can affect the escape process momentarily \citep{2017ApJ...846...31C, 2020A&A...638A..49O, 2022MNRAS.509.5858H}.
\end{enumerate}

\section{Conclusions}
During the Focus Meeting 5 (``Beyond the Goldilocks Zone: the Effect of Stellar Magnetic Activity on Exoplanet Habitability''), we heard about the importance of thinking beyond the goldilocks zones. In this (very brief!) review, I discussed that stellar winds, coronal mass ejections, magnetic activity, stellar high-energy radiation, and planetary magnetic fields are additional important aspects to consider when thinking about planetary habitability. I highlighted 3 key points in my talk and in this article: 1) do planetary magnetic field act as a shield to prevent the loss of planetary atmospheres or a channel that could lead to enhanced escape? 2) Earth-like planets around M dwarfs might not have Earth-like magnetospheres. 3) Atmospheric escape and the evolution of planets depends on the X-ray and EUV history of the host star.

\section*{Acknowledgements}
I thank the organisers of Focus Meeting 5 for hosting a very exciting meeting and a special thanks to the General Assembly local organisation, for their huge efforts in making the whole event happen during very difficult pandemic times. AAV has received funding from the European Research Council (ERC) under the European Union's Horizon 2020 research and innovation programme (grant agreement No 817540, ASTROFLOW).

\def\apj{{ApJ}}    
\def\nat{{Nature}}    
\def\jgr{{JGR}}    
\def\apjl{{ApJ Letters}}    
\def\aap{{A\&A}}   
\def\mnras{{MNRAS}}
\def\aj{{AJ}}
\def\araa{{Annual Review of Astronomy and Astrophysics}}

\let\mnrasl=\mnras

\end{document}